\documentclass[conference]{IEEEtran}
\IEEEoverridecommandlockouts
\usepackage{cite}
\usepackage{amsmath,amssymb,amsfonts}
\usepackage{graphicx}
\usepackage{textcomp,bm}
\usepackage{mathrsfs}
\usepackage{xcolor}
\usepackage{algorithm}%
\usepackage{algorithmicx}%
\usepackage{algpseudocode}%
\usepackage{url,soul,color,balance}
\usepackage{placeins, url}
\usepackage{newtxtext} 
\usepackage{newtxmath} 
\usepackage[T1]{fontenc}

\makeatletter
\let\MYcaption\@makecaption
\makeatother

\usepackage[font=footnotesize]{subcaption}

\makeatletter
\let\@makecaption\MYcaption
\makeatother

\begin{document}

\title{Physical Layer Authentication With Colored RIS \\
in Visible Light Communications\\

\thanks{This work has been done when Besra Çetindere Vela and Serkan Vela were visiting the Dept. of Info. Eng. in Padova, within the Erasmus mobility Program.}
}

\author{
\IEEEauthorblockN{Besra Çetindere Vela\IEEEauthorrefmark{1} and Serkan Vela\IEEEauthorrefmark{1}}
\IEEEauthorblockA{Dept. of Electrical and Electronics Eng.\\
Karadeniz Technical University, 
Trabzon, Turkey \\
\{besracetindere,  serkanvela\}@ktu.edu.tr}
\and
\IEEEauthorblockN{Stefano Tomasin}
\IEEEauthorblockA{Dept. of Information Eng. \\
University of Padova, Italy\\
stefano.tomasin@unipd.it}
\thanks{\IEEEauthorrefmark{1}Besra Çetindere Vela and Serkan Vela contributed equally to this work.}
}

\renewcommand{\arraystretch}{1.3}

\maketitle

\begin{abstract}
We study a visible light communication (VLC) system that employs a colored reconfigurable intelligent surface (CRIS) based on dichroic mirrors that reflect light at tunable frequencies. A verifier can use the CRIS to authenticate transmissions by comparing received multicolor power profiles with expected patterns. Four CRIS configuration strategies are evaluated: a deterministic cyclic pattern, static random reflectance, dynamic random reflectance, and dynamic random permutation of fixed profiles. Randomized configurations, especially dynamic ones, achieve superior authentication, enabling a novel challenge-response physical-layer authentication scheme over CRIS.
\end{abstract}


\section{Introduction} 
Reconfigurable Intelligent Surfaces (RIS) have emerged as a promising technology to enhance visible light communication (VLC). RISs can manipulate optical signals through reflective or refractive elements, widening the angle of reception and mitigating signal loss \cite{10.1109/mwc.001.2000365} and extending coverage~\cite{10.3390/s24020337}. They also increase the achieved data rate by providing stronger signals at the receiver~\cite{10.1109/jphot.2022.3211730}. This capability helps address inherent VLC limitations such as range constraints and interference vulnerability \cite{aboagye2022ris}. They have been proven particularly useful in indoor scenarios~\cite{maraqa2023optimized}. 

Security is another critical area where RIS provides benefits. By optimizing interference management and strategically reflecting signals, RIS can bolster confidentiality in VLC, minimizing eavesdropping opportunities and enhancing secrecy capacity \cite{meghraoui2024secure, khoshafa2024ris}. RISs also enable novel physical-layer authentication (PLA) mechanisms \cite{10623110}. The unique channel is obtained when light passes through an RIS.

In this work, we investigate the application of colored RIS (CRIS) for Physical Layer Authentication (PLA) in VVLC systems. The integration of colored RIS allows for a more granular characterization of the propagation channel, as it leverages distinct optical wavelengths (colors). Moreover, the tunable nature of the CRIS empowers the verifier to actively manipulate the optical environment actively, facilitating the design of advanced authentication mechanisms. In particular, following the framework introduced in \cite{tomasin22}, the RIS enables the implementation of a {\em challenge-response} (CR) PLA scheme consisting of two phases. In the first phase, the verifier collects estimates of the end-to-end channel through the RIS across multiple configurations. In the second phase, a randomly selected RIS configuration is employed upon message reception, and the verifier authenticates the message by verifying that the observed channel matches the expected one associated with the chosen configuration.

 The remainder of this paper is organized as follows: Section II details the system model, Section III describes the CRIS implementation, Section IV explains the PLA approaches, and Section V presents the numerical results comparing the different PLA mechanisms. Finally, Section VI concludes the paper.

\begin{figure}
\centering
  \includegraphics[width=0.9\linewidth]{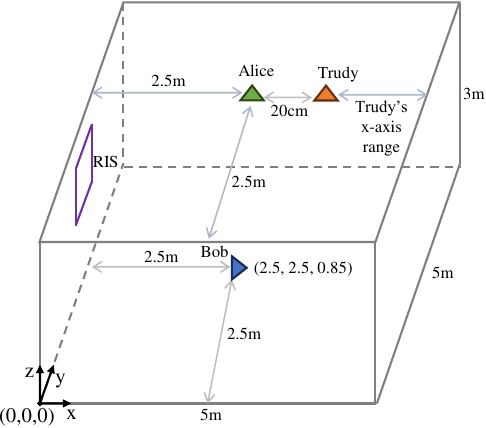}
  \caption{Considered VLC security scenario.}
\hfill
  \label{fig:setup}
\end{figure}

\section{System Model}

\begin{figure}
\centering
  \includegraphics[height=6cm, width=0.7\linewidth]{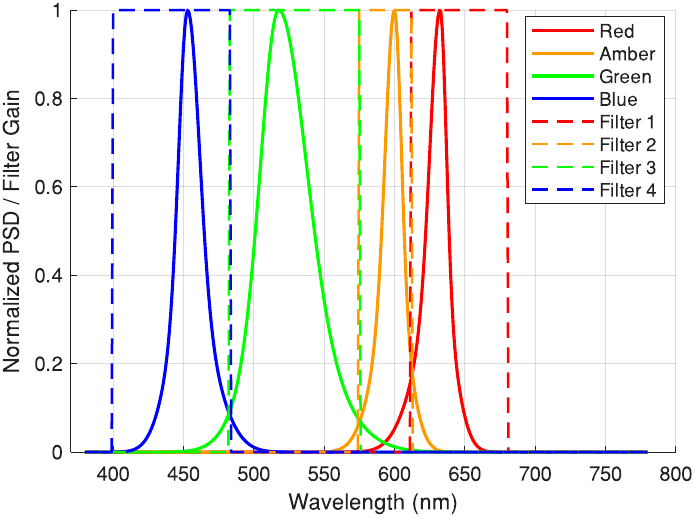}
  \caption{Relative power spectra of the LUXEON C Color\textsuperscript{\textregistered} QLED.}
  \label{fig:psd}
\hfill
\end{figure}

We consider the scenario described in Fig.~\ref{fig:setup},  wherein agent Alice transmits messages through a VLC system to agent Bob. An attacker, Trudy, also transmits messages to Bob {\em impersonating} Alice, i.e., forging the messages so that they appear as sent by Alice. Bob aims at checking if received messages have been truly transmitted by Alice, through an {\em authentication} mechanism. 

Authentication occurs at the physical layer by leveraging the unique properties of received optical signals. Both Alice and the adversary Trudy use identical four-color QLEDs emitting red, amber, green, and blue light. Bob, the verifier, employs four photodetectors (PDs), each paired with an ideal optical filter for one of the four wavelengths. As shown in Fig.~\ref{fig:setup}, Alice and Trudy are mounted on the ceiling, directing their QLEDs downward, while Bob is positioned below, facing a wall-mounted colored RIS. There is no direct line-of-sight (LoS) path between the transmitters and Bob. The dichroic mirrors in the CRIS are precisely angled to reflect light toward Bob’s PDs, ensuring high signal quality and improved resistance to eavesdropping.

\subsection{LED Model} 

Various models can represent the QLED's power spectral density (PSD). The H model outlined in \cite{he2010model} closely resembles the findings of the experimental studies. The parameters of the H model, as presented in Table \ref{tab:1}, are chosen to accurately simulate the PSD of the commercially available LUXEON C Color LED\textsuperscript{\textregistered} based on the data provided in its datasheet \cite{luxeon_datasheet}. These parameters are then utilized to calculate the shape of the PSD of the QLED. In particular, the left and right half spectral widths are denoted by $\Delta\lambda_1$ and $\Delta\lambda_2$, respectively. Denoting the peak wavelength of the PSD as $\lambda_p$ and defining the functions
\begin{equation}
g(\lambda, \lambda_p, \Delta\lambda) = \exp\left[-\frac{(\lambda - \lambda_p)^2}{\Delta\lambda^2}\right],
\label{eqn:6}
\end{equation}
\begin{equation}
\Delta\lambda = 
\begin{cases}
\Delta\lambda_1, & \lambda < \lambda_p \\
\Delta\lambda_2, & \lambda \geq \lambda_p
\end{cases},
\label{eqn:7}
\end{equation}
The power spectral density (PSD) is given by
\begin{equation}
\Phi_c(\lambda) = \frac{g(\lambda, \lambda_p, \Delta\lambda) + k_1 g^{k_2}(\lambda, \lambda_p, \Delta\lambda)}{1 + k_1}.
\label{eqn:8}
\end{equation}
The relative energy emitted in spectral channel $c \in \{R,\,A,\,G,\,B\}$ is
\begin{equation}
\mathcal{E}_c = \int_{\lambda_{c,\ell}}^{\lambda_{c,u}} \Phi_c(\lambda) \, d\lambda,
\label{eq:spec_power}
\end{equation}
where $\mathcal{E}_c$ denotes the integrated PSD over the channel-specific wavelength range.

\begin{table} 
\centering
\caption{H Model Parameter Values of The Considered QLED \cite{luxeon_datasheet}.}
\begin{tabular}{|c c c c c|}
\hline
\textbf{Parameter} & \textbf{Red} & \textbf{Green} & \textbf{Blue} & \textbf{Amber} \\
\hline
$\lambda_p$ & 632.5 & 600 & 517.7 & 453 \\
$\Delta\lambda_1$ & 23.84 & 19.66 & 29.38 & 18.99 \\
$\Delta\lambda_2$ & 14.74 & 14.97 & 45.21 & 25.5 \\
$k_1$ & 2 & 2 & 2 & 2 \\
$k_2$ & 6 & 5 & 3 & 5 \\
\hline
\end{tabular}
\vspace{-2ex}
\label{tab:1}
\end{table}

\begin{figure}
\centering
\begin{tabular}{cc}
  \includegraphics[height=3.3cm, width=0.6\linewidth]{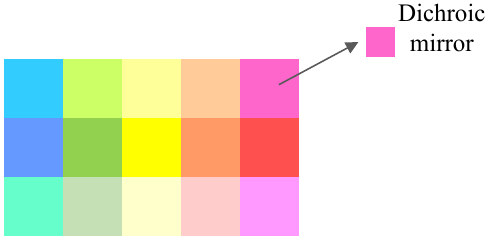} &
    \includegraphics[height=3.7cm, width=0.36\linewidth]{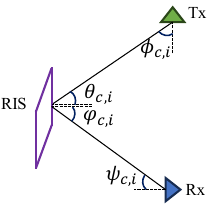}\\
    a) & b) 
    \end{tabular}
  \caption{a) Illustration of the CRIS of $5 \times 3$ elements, with a specific configuration (described by the color of each element), where each element is configured with different reflective parameters. b) 
Angle notation for the considered scenario.}
  \label{fig:CRIS}
\hfill
\end{figure}
\subsection{Power Model}

The total optical power received at Bob from the transmitters, denoted by $x \in \{\text{A},\,\text{T}\}$ (Alice and Trudy, respectively), for each color channel $c \in \{R,\,A,\,G,\,B\}$ (representing the red, amber, green, and blue channels, respectively) can be expressed as 
\begin{equation}
    P_c = E_{x,c} + W_c,
\end{equation}
where $E_{x,c}$ is the noise-free received power at Bob for the color channel $c$ from transmitter $x$, and $W_c$ denotes the corresponding zero-mean additive white Gaussian noise (AWGN) power with variance $\sigma_c^2$ at Bob's PD. The received signal component is given by:
\begin{equation}\label{chmod}
    E_{x,c} = h_{x,c}^{2} \mathcal{P}_{x,c},
\end{equation}
where $h_{x,c}$ represents the channel gain between transmitter $x$ and Bob for the color channel $c$, and $\mathcal{P}_{x,c}$ denotes the transmitted LED power from transmitter $x$ for the corresponding color channel. We have assumed that these powers are the same for each color.




\section{Colored RIS with Dichroic Mirrors} 

The CRIS is an innovative structure utilized in VLC systems, comprising an array of dichroic mirrors (elements), i.e., mirrors capable of selectively reflecting specific wavelengths of visible light. The reflective properties of each element are {\em tunable}, thus can be modified by the CRIS controller, which in our scenario is Bob. A specific setting of the reflective properties of each element is denoted as a {\em CRIS configuration}.

An illustration of an example of CRIS and a specific configuration is provided in Fig.~\ref{fig:CRIS}: here, each dichroic mirror exhibits a distinct reflectance profile as a function of wavelength $\lambda$, i.e., each element reflects a different color when illuminated with a white light. 

The dichroic mirrors of the CRIS are precisely engineered using multilayer dielectric coatings, allowing accurate control over their spectral reflectivity and transmission properties. Furthermore, by incorporating tunable (adjustable) optical coatings or mechanically reconfigurable elements, CRIS elements can dynamically modify their reflective properties, adapting to different operational scenarios. This feature allows CRIS to tune spectral reflection, directing specific color channels—such as red, amber, green, or blue—to desired receiver positions while suppressing or attenuating others \cite{lee2014color}.

The reflective properties of element $i$ of the CRIS are defined by the reflectivity coefficients for the four colors, $\rho_{c,i}$, $c \in \{R,\,A,\,G,\,B\}$. Note that a specific color as shown in Fig.~\ref{fig:CRIS}.a is obtained by linearly the power spectra of the colors in Fig.~\ref{fig:psd} with combining coefficients $\rho_{c,i}$, $c \in \{R,\,A,\,G,\,B\}$.

A specific CRIS configuration is defined by the set 
\begin{equation}
\mathcal C = \{\rho_{c,i}, \forall c \in \{R,\,A,\,G,\,B\}, \forall i \in [1,N]   \}
\end{equation}
where $N$ denotes the number of CRIS reflective elements.

\subsection{Multi-color CRIS Reflected VLC Channels}


In this study, we adopt the well-known VLC channel model of~\cite{lee2011indoor} to characterize the multicolor VLC channel between Alice (or Trudy), the CRIS, and Bob. 

In particular, the total channel gain reflected through CRIS for color $c$ is $h_{x,c}$ (see \eqref{chmod}). For a simpler notation, we drop the transmitter index to $x$. 

For each color channel $c \in \{R,\,A,\,G,\,B\}$  is given by  
\begin{equation}
    h_c = \sum_{i=1}^{N} h_{c,i}^{(lr)} \, h_{c,i}^{(rp)},
\end{equation}
where $N$ is the number of CRIS elements, $h_{c,i}^{(lr)} \in \mathbb{R}$ denotes the NLoS channel gain from Alice (or Trudy) to the $i$-th RIS element, and $h_{c,i}^{(rp)} \in \mathbb{R}$ is the channel gain from the $i$-th RIS element to Bob's PD.

For the computation of the channel gains, we refer to the angular notation shown in Fig.~\ref{fig:CRIS}.b. The channel gain from Alice or Trudy to the $i$-th RIS element for each color is calculated as follows:
\begin{equation}
    h_{c,i}^{(lr)} = \frac{(m+1) A_{\text{CRIS}} \cos^m\phi_{c, i} \cos\theta_{c, i}}{2 \pi d_{c, i}^2},
\end{equation}
where $m = -1/\log_2 \cos(\theta_{1/2})$ is the Lambertian order of the LED, and $\theta_{1/2}$ represents the half-view angle of the LED. Here, $A_{\text{CRIS}}$ denotes the reflective surface area of each RIS element, and $d_{c,i}$ is the distance between the transmitter and the $i$-th RIS element. The irradiance angle at the transmitter and the incidence angle at the $i$-th RIS element are denoted as $\phi_{c,i}$ and $\theta_{c, i}$, respectively.

The channel gain from the $i$-th CRIS element to Bob's PD for $c$-th color is expressed as:
\begin{equation}
    h_{c,i}^{(rp)} = \frac{ A_{\text{PD}} \cos\varphi_{c, i}\cos\psi_{c, i}}{\pi d_{c, i}^2} G R \rho_{c, i},
\end{equation}
where $A_{\text{PD}}$ is the effective area of Bob's PD, $d_{c,i}$ denotes the distance between the RIS element and Bob, and $\rho_{c, i}$ is the reflectivity coefficient of the RIS element specific to the color channel $c$. Furthermore, $G$ represents the gain of Bob's optical concentrator, given as $G = n^2/\sin^2\theta_{\text{fov}}$, where $n$ is the refractive index of the PD lens, and $\theta_{\text{fov}}$ is the field-of-view angle of the PD. The angles $\varphi_{c,i}$ and $\psi_{c,i}$ represent the irradiance angle at the $i$-th RIS element and the incidence angle at Bob's PD, respectively. 

Note that the reflectivity coefficient $\rho_{c, i}$ depends on the design of the dichroic mirror for each color channel $c$. For example, an idealized mirror might reflect only one color channel ($\rho_{c,i} = 1$ for the target color, 0 otherwise), while a practical device achieves reflectivity $\rho_{c,i} \in [0,1]$.

\section{Physical Layer Authentication}

The PLA mechanism is implemented by Bob, who must decide whether a received message is from Alice or not. The attacker device Trudy, on the other hand, aims to send messages to Bob that he believes to be from Alice. Trudy is identical to Alice, using the same QLED device with equivalent emission spectrum, power levels, and directional orientation. However, Trudy is in a different position with respect to Alice, so the optical signal received by Bob (also through the CRIS) will be different.

We consider two PLA mechanisms with CRIS: the first, denoted single-configuration PLA (SC-PLA), uses a single CRIS configuration, while the latter, denoted challenge-response PLA (CR-PLA), exploits a random behavior of the CRIS. In any case, we assume that Bob does not know anything about Trudy, in particular, he does not know her channel to the CRIS. Thus, the authentication test is based uniquely on the knowledge of the Alice-Bob channel.

\subsection{Single-Configuration PLA}

In the single-configuration PLA \cite{6204019}, PLA includes two phases: a) the identification association (IA), and b) the identification verification (IV) phase. The IA is performed whenever the channel significantly changes, and here we assume a static scenario, thus, this phase is performed only once. In this phase, the communication between Alice and Bob is authenticated (e.g., using a cryptographic approach), Alice transmits and Bob estimates the received energy for the various colors; let $\hat{E}_{A,c}$, $c \in \{R,\,A,\,G,\,B\}$, be the estimated energy in this phase, which is an AWGN-corrupted version of $E_{A,c}$. 

Instead, the IV phase is performed by Bob when he receives a new message: Bob again estimates the received energies for different colors and compares them with those estimated in the first phase. Let the received powers be $P_c$, $c \in \{R,\,A,\,G,\,B\}$ for each color. Considering that its estimates are affected by AWGN with independent components per color, the likelihood test for authentication is given by \cite{6204019}
\begin{equation} 
L(\bm {P}) = \sum_c(P_c - \hat{E}_{A,c})^2 \begin{cases}
< \gamma & {\rm authentic}, \\
\geq \gamma & {\rm not\ authentic}, \\
\end{cases}
\label{GLRT}
\end{equation}
where $\bm{P} = [P_R, P_A, P_ G, P_B]$ and $\gamma$ is a suitable threshold. In particular, for a value of $L(\bm{P})$ larger than $\gamma$ an attack is declared, while for a value of $L(\bm{P})$  smaller than $\gamma$ the message is accepted as authentic. 

Two key performance metrics are the probability of false alarm ($P_{\rm FA}$), where Bob incorrectly rejects Alice's legitimate transmission ($P_{\rm FA} = {\mathbb P}[L(\bm {P}) > \gamma| \mbox{Alice transmits}]$), and the probability of misdetection ($P_{\rm MD}$), where Bob incorrectly accepts Trudy's impersonation attempt $P_{\rm MD} = {\mathbb P}[L(\bm {P}) \leq \gamma| \mbox{Trudy transmits}]$.

Two options were considered for selecting the static configuration in the single-configuration PLA, referred to as Scenarios 1 and 2.

\paragraph*{Scenario 1 -- Fixed Reflectance} In this scenario, the RIS mirrors employ the predetermined, cyclic reflectivity pattern utilizing four specific reflectance profiles (detailed in Section VI.A).

\paragraph*{Scenario 2 --  Static Random Reflectance} In this configuration, the mirror reflectance coefficients $\rho_{c,i}$ are independent realizations of uniform random variables in  [0, 1] for each color \(c\) and mirror \(i\).

\subsection{Challenge-Response PLA}

In the CR-PLA mechanism, a random CRIS configuration is used at each message transmission. In particular, we still have the two phases of a) the identification association, and b) the identification verification phase. 

However, in CR-PLA \cite{tomasin22} in the first IA phase (authenticated at higher layers), several CRIS configurations are explored, and Bob estimates the received power for each of them. The purpose of this phase is to enable Bob to predict the received energies for {\em any} CRIS configuration. 

Indeed, in the IV phase, Bob {\em chooses randomly} a CRIS configuration,  estimates the received energies for various colors, and compares them with those predicted from the estimates obtained in the first phase. Let now $\hat{E}_c$ be the predicted energies for the currently selected configuration. Then, again, the LT \eqref{GLRT} is performed to decide on the authenticity of the received message. 

Note that the advantage of CR-PLA is its high resistance against multiple repeated attacks, since the randomness introduced in each transmission, which is not known to Trudy, makes the attacks from Trudy harder. 

\paragraph*{Scenario 3 -- Dynamic Random Reflectance} For CR-PLA, at each transmission, the reflective coefficients of the elements are randomly changed, with the same procedure as Scenario 2. In particular, each reflective parameter $\rho_{c,i}$ is the realization of an independent uniform random variable in the interval $[0,1]$.  

\paragraph*{Scenario 4 -- Random Reflectance Permutation} This scenario utilizes the same set of four specific color reflectance profiles defined in Scenario 1. However, instead of assigning them in a fixed cyclic order, these four patterns to the $N$ CRIS elements are randomly permuted at each transmission instance. The number of distinct reflectance profiles used is limited to 4, unlike the continuous range in Scenario 3, with randomness arising solely from the permutation of their positions.

\begin{table}
\centering
\caption{Simulation Parameters}
\begin{tabular}{|l l|}
\hline
\textbf{Parameter} & \textbf{Value} \\
\hline
CRIS element size ($A_{\text{CRIS}}$) & $10\,\text{cm}\times10\,\text{cm}$ \\
Room dimensions & $5\,\text{m}\times5\,\text{m}\times3\,\text{m}$ \\
CRIS central coordinates & $[0,\,2.5,\,1.5]$ \\
Alice coordinates & $[2.5,\,2.5,\,3]$ \\
Bob coordinates & $[2.5,\,2.5,\,0.85]$ \\
LED half-viewing angle ($\theta_{1/2}$) & $47.5^\circ$ \\
Optical filter lower limits ($\lambda_{c,\ell}$) & $[612,\,575,\,483,\,400]\,\text{nm}$ \\
Optical filter upper limits ($\lambda_{c,u}$) & $[680,\,612,\,575,\,483]\,\text{nm}$ \\
Optical filter gain & $1$ \\
PD Field of View ($\theta_{\text{fov}}$) & $120^\circ$ \\
PD refractive index ($n$) & $1.5$ \\
PD effective area ($A_{\text{PD}}$) & $1\,\text{cm}^2$ \\
PD responsivity ($R$) & $0.54\,\text{A/W}$ \\
Transmitted optical power per color ($\mathcal{P}_{x,c}$) & $25\,\text{W}$ \\
\hline
\end{tabular}
\vspace{-2ex}
\label{tab:simulation_parameters}
\end{table}

\subsection{Attacker Model and Strategy}
\label{attacksec}

In this study, we consider two types of attack strategies with varying levels of attacker capability. In the simpler model, Trudy transmits directly from her location without applying any signal processing or precoding. This allows us to evaluate the inherent robustness of the proposed authentication system under passive impersonation, where the difference in spatial location alone serves as the basis for authentication.

\paragraph*{LoS Attack} As a more advanced strategy, we consider a Line-of-Sight (LoS) impersonation attack, in which Trudy bypasses the RIS entirely and transmits directly toward Bob. We assume Trudy possesses partial knowledge of the system: she knows the \emph{total average noise-free received power} expected at Bob from Alice, aggregated across all color channels and averaged over the possible CRIS configuration, $\bar{E}_A = \sum_{c} \mathbb{E}_{\mathcal{C}}[E_{A,c}(\mathcal{C})]$, where the expectation is taken over the random CRIS configurations used in CR-PLA. Trudy, however, does \textit{not} have access to the individual color-specific components of this power. Then, she selects transmit power for color $c$ such that 
\begin{equation}
    \sum_c E_{T,c}^{\text{LoS}} = \bar{E}_A,
\end{equation}
where $E_{T,c}^{\text{LoS}} = \mathcal{P}_{T,c}^{\text{LoS}} (h_{T,c}^{\text{LoS}})^2$, and $h_{T,c}^{\text{LoS}}$ is the LoS channel gain from Trudy to Bob for color $c$.   Under this impersonation attempt, the total optical power received at Bob on each color channel becomes
\begin{equation}
    P_c^{\text{attack}} = E_{T,c}^{\text{LoS}} + W_c,
\end{equation}
where $W_c$ is additive white Gaussian noise with variance $\sigma_c^2$. 




\section{Numerical Results}

We simulated the VLC system in a $5\times5\times3\,\text{m}^3$ room. Alice is at $[2.5, 2.5, 3]$~m, Bob at $[2.5, 2.5, 0.85]$~m (facing the CRIS), and the CRIS center is at $[0, 2.5, 1.5]$~m. Key parameters are listed in Table~\ref{tab:simulation_parameters}. We evaluate CRIS sizes $N = 960$ and $1500$, corresponding to grid layouts of $40 \times 24$, and $50 \times 30$. Trudy's passive impersonation location varies along the x-axis: $x_T \in \{2.7, 3.05, 3.4, 3.75, 4.1\}$~m ($y=2.5$~m, $z=3$~m). The LoS attacker is at $[0.1, 2.5, 0.85]$~m, facing Bob.





\begin{figure} 
\centering
  \includegraphics[width=0.9\linewidth]{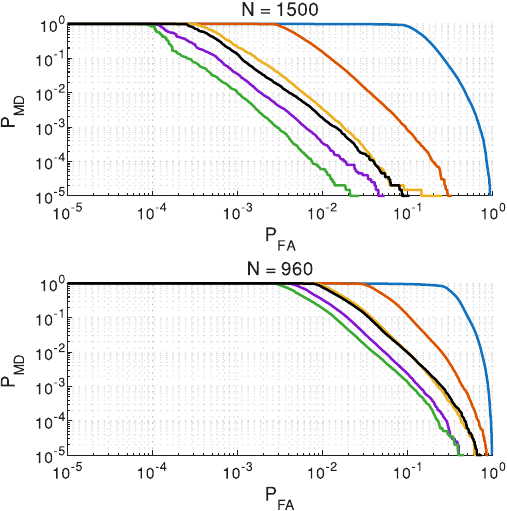}
\caption{Scenario~1 DET curves for varying $N$. Curve colors indicate Trudy’s $x$-position (Section~VI). Black: LoS attack.}

  \label{fig:Case_1_Fig1}
\end{figure}

\begin{figure} 
\centering
  \includegraphics[width=0.9\linewidth]{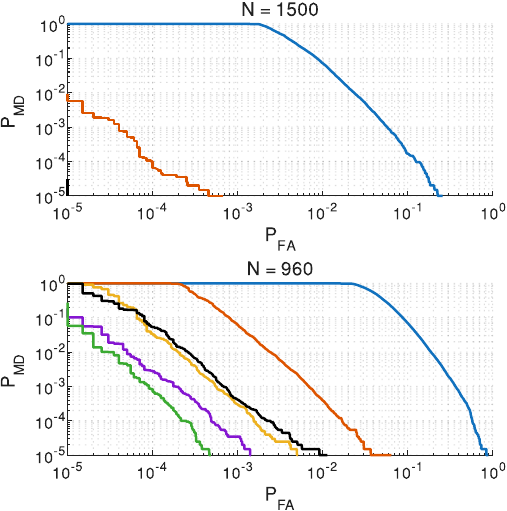}
\caption{Scenario~2 DET curves for varying $N$. Curve colors indicate Trudy’s $x$-position (Section~VI). Black: LoS attack.}

  \label{fig:Case_2_Fig1}
\end{figure}

 \begin{figure} 
 \centering
   \includegraphics[width=0.9\linewidth]{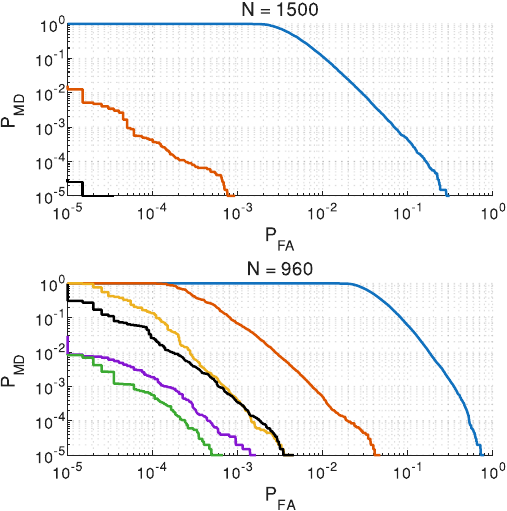}
\caption{Scenario~3 DET curves for varying $N$. Curve colors indicate Trudy’s $x$-position (Section~VI). Black: LoS attack.}

   \label{fig:Case_3_Fig1}
 \end{figure}

  \begin{figure} 
  \centering
\includegraphics[width=0.9\linewidth]{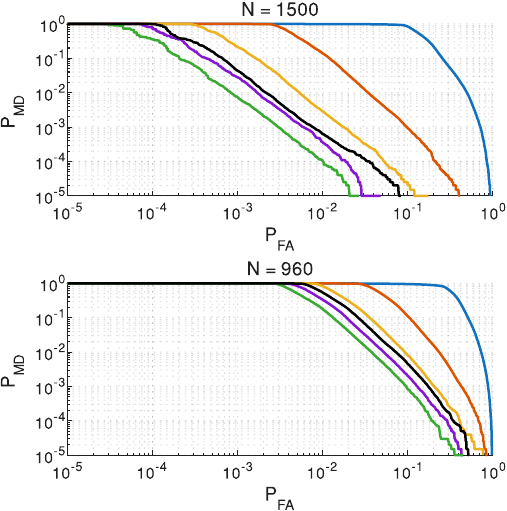}
\caption{Scenario~4 DET curves for varying $N$. Curve colors indicate Trudy’s $x$-position (Section~VI). Black: LoS attack.}

   \label{fig:Case_4_Fig1}
 \end{figure}

To ensure a fair comparison across all CRIS reflectance configurations, a fixed noise variance is used in all simulation scenarios, computed with reference to the first iteration of Scenario~1 (1500 mirrors) as
\begin{equation}
    \sigma^2 = \frac{\sum_{c=1}^{4} h_{A,c}^{2}}{10^{\mathrm{SNR}_{\mathrm{dB}}/10}},
\end{equation}
where $h_{A,c}$ is the CRIS channel gain between Alice and Bob for color channel $c \in \{R, A, G, B\}$, and ${\rm SNR}_{\rm dB} = 10$~dB is the total received SNR across all channels. 

We evaluate authentication performance using the probabilities $P_{\text{FA}}$ and $P_{\text{MD}}$, plotting the detection error tradeoff (DET) curve by varying the decision threshold $\gamma$. The probabilities are estimated using Monte Carlo simulations with 20,000 noise realizations per configuration (transmitter, CRIS size, Trudy's position, and reflectance scenario).

\subsection{Performance Comparison}


We evaluate five impersonation distances for Trudy, increasing her x-axis separation from Alice; these are shown in the DET plots of Fig.~\ref{fig:Case_1_Fig1}–\ref{fig:Case_4_Fig1}. For the LoS attack of Section~\ref{attacksec}, Trudy transmits from $[0.1,\,2.5,\,0.85]$ facing Bob.

Performance details for each scenario are presented below:

\subsubsection*{Scenario~1 -- Fixed Reflectance}
This SC-PLA setup applies a cyclic pattern of four predefined reflectance vectors across CRIS elements, each favoring a different color channel: $[1, 0.2, 0, 0]$ (R), $[0.2, 1, 0.2, 0]$ (A), $[0, 0.2, 1, 0.2]$ (G), and $[0, 0, 0.2, 1]$ (B). As shown in Fig.~\ref{fig:Case_1_Fig1}, this static structure results in weaker authentication, with shallow DET curves, especially under LoS attacks or when Trudy is close to Alice.

\subsubsection*{Scenario~2 -- Static Random Reflectance}
Here, $\rho_{c,i}$ values are independently sampled from $\mathcal{U}[0,1]$ and fixed for each CRIS element and channel. This static randomness improves performance over Scenario~1, yielding steeper DET curves of Fig.~\ref{fig:Case_2_Fig1}, with larger gains as $N$ increases.

\subsubsection*{Scenario~3 -- Dynamic Random Reflectance}
This scenario is similar to Scenario~2, but with $\rho_{c,i}$ re-randomized per transmission. This dynamic randomness delivers the best results overall, Fig.~\ref{fig:Case_3_Fig1}, enhancing channel unpredictability and robustness, especially against LoS attacks.

\subsubsection*{Scenario~4 -- Random Reflectance Permutation}
This CR-PLA scenario uses the same four vectors from Scenario~1, but randomly permuted across CRIS elements at each transmission. While better than Scenario~1, it underperforms compared to Scenarios~2 and~3 due to limited reflectance diversity Fig.~\ref{fig:Case_4_Fig1}.

Here are the key insights from the results:

\begin{itemize}
    \item Scenarios~1 and~4 show nearly identical, weaker performance—permutation alone does not compensate for limited pattern diversity.
    \item Scenario~2 introduces effective diversity, outperforming Scenarios~1 and~4, especially as $N$ grows.
    \item Scenario~3 achieves the strongest results, with dynamic reflectance yielding the steepest DET curves across all conditions.
    \item Larger $N$ and increased Alice–Trudy separation both enhance authentication by improving channel distinguishability.
\end{itemize}

In summary, incorporating randomness, particularly dynamic, per-color reflectance variation, substantially strengthens VLC authentication. Scenario~3 provides the highest resilience, followed by Scenario~2. Limited pattern diversity, as in Scenario~4, offers only marginal improvement over static designs.

\balance
\section{Conclusion}


Simulation results show that CRIS randomization greatly enhances authentication. Scenario~3 (dynamic random reflectance) consistently outperforms others, offering strong resistance to both near-field and sophisticated LoS attackers. Scenario~2 (static random) achieves similar results with slightly less robustness. In contrast, Scenarios~1 and~4 (fixed and permuted patterns) perform significantly worse and yield nearly identical outcomes.


In summary, dynamically reconfigurable, color-selective CRIS designs significantly bolster VLC physical-layer security. Future work will focus on adaptive CRIS control, mobility resilience, and experimental validation.
\balance


\bibliographystyle{IEEEtran}
\bibliography{ref.bib}

\end{document}